\documentclass[twocolumn,reprint,amsmath,amssymb, aps,prl]{revtex4-2}
\usepackage[T1]{fontenc}
\usepackage[latin9]{inputenc}
\usepackage{color}
\usepackage{float}

\makeatletter
\usepackage{amsfonts,amssymb}
\usepackage{latexsym}
\usepackage{epsfig}
\usepackage{graphicx}
\usepackage{tikz}
\definecolor{greatblue}{RGB}{40,120,181}
\definecolor{greatred}{RGB}{200,36,35}
\usepackage[colorlinks,linkcolor=greatblue,anchorcolor=blue,citecolor=greatred]{hyperref}

\usepackage{graphicx}% Include figure files
\usepackage{dcolumn}% Align table columns on decimal point
\usepackage{bm}% bold math

\usepackage{upgreek}
\usepackage{mathrsfs}
\usepackage{mathbbold}
\usepackage{bbm}
\usepackage{youngtab}
\usepackage{slashed}

\newcommand{\ii}{{\rm i}}

\newcommand{\oI}{{\rm I}}
\newcommand{\oII}{{\rm II}}

\newcommand{\Vy}{{\bf y}}
\newcommand{\Vtheta}{{\uptheta}}
\newcommand{\Vxi}{{\bm\xi}}
\newcommand{\Vw}{{\bf w}}

\newcommand{\hVw}{{\hat{\Vw}}}

\newcommand{\cM}{{\mathcal M}}
\newcommand{\rd}{{\rm d}}
\newcommand{\rD}{{\rm D}}

\newcommand{\hA}{{\hat A}}
\newcommand{\hB}{{\hat B}}
\newcommand{\hC}{{\hat C}}

\newcommand{\ha}{{\hat a}}
\newcommand{\hb}{{\hat b}}

\newcommand{\halpha}{{\hat\alpha}}
\newcommand{\hbeta}{{\hat\beta}}
\newcommand{\hgamma}{{\hat\gamma}}
\newcommand{\heta}{{\hat\eta}}
\newcommand{\hblt}{{\hat\bullet}}

\newcommand{\ShA}{{\hat{\mathbb A}}}
\newcommand{\ShB}{{\hat{\mathbb B}}}
\newcommand{\ShC}{{\hat{\mathbb C}}}
\newcommand{\SC}{{\mathbb C}}
\newcommand{\SI}{{\mathbb I}}
\newcommand{\Shalpha}{{\hat\bbalpha}}
\newcommand{\Shbeta}{{\hat\bbbeta}}

\newcommand{\Stheta}{{\bbtheta}}
\newcommand{\Seta}{{\bbeta}}
\newcommand{\STheta}{{\mathbbold\Theta}}
\newcommand{\Sy}{{\mathbbm y}}
\newcommand{\SY}{{\mathbbm Y}}
\newcommand{\SA}{{\mathbbm A}}
\newcommand{\SD}{{\mathbbm D}}
\newcommand{\SF}{{\mathbbm F}}
\newcommand{\Su}{{\mathbbm u}}
\newcommand{\ff}{{\mathfrak{f}}}
\newcommand{\SE}{{\mathbbm E}}

\newcommand{\Bu}{{u}}

\newcommand{\BF}{{F}}
\newcommand{\BSF}{{\mathcal F}}
\newcommand{\SSF}{{\mathscr F}}
\newcommand{\STr}{{\rm STr}}

\makeatother

\usepackage[english]{babel}

\begin{document}

\title{On AdS$_{4}$ superspace and supergravity}

\author{Chen-Xu Han$^{1}$}
%\email{???}
%\affiliation{Institute of Modern Physics, Northwest University, Xi'an 710127, China}

\author{Zhao-Long Wang$^{1,2,3}$}
\email{zlwang@nwu.edu.cn}

\author{Yi Yan$^{1}$}
%\email{???}
%\affiliation{Institute of Modern Physics, Northwest University, Xi'an 710127, China}

\affiliation{$^1$Institute of Modern Physics, Northwest University, Xi'an 710127, China
\\$^2$Peng Huanwu Center for Fundamental Theory, Xi'an 710127, China
\\$^3$Shaanxi Key Laboratory for Theoretical Physics Frontiers, Xi'an 710127, China}

%\date{\today}

\begin{abstract}
In the $N=1$ superspace, AdS$_4$ supersymmetry is realized as the non-linear super coordinate transformations. The fermionic coordinates form a faithful non-linear representation of supersymmetry on their own. By introducing an auxiliary scalar coordinate, this representation is reformulated as a 5-dimensional linear representation, i.e., the superspinor representation. New linear representations are constructed by tensor products of multiple superspinors. Especially, the superspace bosonic coordinates are embedded in the supervector representation, which is the traceless symmetric part of the bi-superspinor representation. Based on these linear representations, the $N=1$ AdS$_4$ supergravity action can be reproduced in a manifestly supersymmetry-covariant way.
\end{abstract}

\maketitle
\onecolumngrid

\section{Introduction}
Einstein's theory of gravity is a non-Abelian gauge theory of a spin-2 particle. The gauge symmetries of the theory can be naturally identified with the global spacetime symmetries of its most symmetric vacua. In the presence of a negative cosmological constant $\Lambda=-\frac{d(d-1)}{2\ell^2}$, the maximal symmetric vacuum is AdS$_{d+1}$, and thus the corresponding gauge group is $SO(2,d)$. For $D=4$, the $SO(2,3)$ gauge symmetry of Einstein gravity was manifest in the formulation by MacDowell and Mansouri \cite{MacDowell:1977jt,Mansouri:1977ej}. An improved version was presented by Stelle and West in \cite{Stelle:1979aj}, where the action is constructed in a way that is manifestly covariant under the $SO(2,3)$ gauge transformations.  An alternative formula of manifestly gauge covariant construction for Einstein gravity was introduced in \cite{Wilczek:1998ea}, and its generalization to arbitrary dimensions was presented in \cite{Wang:2018tyq}. The basic idea in \cite{MacDowell:1977jt,Mansouri:1977ej,Stelle:1979aj,Wilczek:1998ea,Wang:2018tyq} is to unify the $SO(1,d)$ spin connection $\omega^{\ha}{}_{\hb}$ and the vielbein $e^{\ha}$ as different components of the $SO(2,d)$ gauge connection 1-form
\begin{eqnarray}
A^{\halpha}{}_{\hbeta}=\left(
                                 \begin{array}{cc}
                                   A^{\hblt}{}_{\hblt} & A^{\hblt}{}_{\hb}\\
                                   A^{\ha}{}_{\hblt} & A^{\ha}{}_{\hb} \\
                                 \end{array}
                               \right)=\left(
                                 \begin{array}{cc}
                                   0 & -\ell^{-1}e_{\hb}\\
                                   \ell^{-1}e^{\ha} & \omega^{\ha}{}_{\hb} \\
                                 \end{array}
                               \right)
\end{eqnarray}
where $\ha,\hb=\hat0,\cdots,\hat d$ are $SO(1,d)$ vector indices and $\halpha,\hbeta=\hblt,\hat0,\cdots,\hat d$ are $SO(2,d)$ vector indices. In \cite{Stelle:1979aj,Wilczek:1998ea,Wang:2018tyq}, an auxiliary field $Y^{\halpha}(x)$
was introduced in order to manifest the gauge covariance. The auxiliary field satisfies the constraint $\heta_{\halpha\hbeta}Y^{\halpha}Y^{\hbeta}=-\ell^2$ where $\heta={\rm diag}\{-1,-1,1,\cdots,1\}$. By local $SO(2,d)$ rotations, $Y^{\halpha}(x)$ can always be fixed to the Einstein gauge
\begin{eqnarray}
\{Y^{\hblt}(x)=\ell\,,~~~~Y^{\ha}(x)=0\}
\end{eqnarray}
and the residual gauge symmetry becomes $SO(1,d)$.
In the language of fiber bundle, we are dealing with a AdS bundle $E$
\begin{eqnarray}
{\rm AdS}_{d+1} \longrightarrow E \overset{\pi}{\longrightarrow} \cM_{d+1}
%:\{\Vy^{\halpha}\}\to \{\Vy^{\halpha},x^{\mu}\} \overset{\pi}{\to} \{x^{\mu}\}
\end{eqnarray}
upon the $d+1$ dimensional base manifold $\cM_{d+1}$ with coordinates $\{x^{\mu}\}$.
The fiber space is the pure AdS space
\begin{eqnarray}
{\rm AdS}_{d+1}=\frac{SO(2,d)}{SO(1,d)}
\end{eqnarray}
which can be described by the embedding coordinates $\Vy^{\halpha}$.
%\begin{eqnarray}
%\heta_{\halpha\hbeta}\Vy^{\halpha}\Vy^{\hbeta}=-\ell^2\,.
%\end{eqnarray}
The auxiliary field $Y^{\halpha}(x)$ is just a section $\Vy^{\halpha}=Y^{\halpha}(x)$ of the AdS bundle.
In terms of $Y^{\halpha}$ and the curvature 2-form $F^{\halpha}{}_{\hbeta}=\rd A^{\halpha}{}_{\hbeta}+A^{\halpha}{}_{\hgamma}\wedge A^{\hgamma}{}_{\hbeta}$, the action of AdS$_{4}$ Einstein gravity is constructed gauge covariantly \cite{Stelle:1979aj} as
\begin{eqnarray}\label{SW}
S_4=\int_{\cM_{4}} \epsilon_{\halpha_0\cdots\halpha_4}F^{\halpha_0\halpha_1}\wedge F^{\halpha_2\halpha_3}Y^{\halpha_4}\,,
\end{eqnarray}
where $\epsilon_{\halpha_0\cdots\halpha_4}$ is the totally antisymmetric tensor. In higher dimensions, if one generalize the MacDowell-Mansouri-Stelle-West type of action by including more $F$'s,  it will lead to a gravity theory with higher derivative terms. To get the $SO(2,d)$ gauge covariant formulation for Einstein gravity in any dimension, one need to
include the $SO(2,d)$ gauge covariant derivative $\rD Y^{\halpha}=\rd Y^{\halpha}+A^{\halpha}{}_{\hbeta}  Y^{\hbeta}$ as well. Then the alternative action \cite{Wilczek:1998ea,Wang:2018tyq} valid for any dimension is constructed as
\begin{eqnarray}\label{VW}
S_{d+1}=\int_{\cM_{d+1}}  \epsilon_{\hat\alpha_0\cdots \hat\alpha_{d+1}} \left[ F^{\hat\alpha_0\hat\alpha_1}-\frac{2}{(d+1)\ell^2} \rD Y^{\hat\alpha_0}\wedge  \rD Y^{\hat\alpha_1}\right]\wedge  \rD Y^{\hat\alpha_2}\wedge\cdots \wedge  \rD Y^{\hat\alpha_d}Y^{\hat\alpha_{d+1}}\,.
\end{eqnarray}
For $d+1=4$, (\ref{VW}) is equivalent to (\ref{SW}) up to a local total derivative term.

By the same logic, it is natural to seek a manifestly supersymmetry-covariant construction of supergravity, which is the non-abelian gauge theory of supersymmetry. In fact, the MacDowell-Mansouri type of action for N=1 AdS$_4$ supergravity was already presented in the original works \cite{MacDowell:1977jt,Mansouri:1977ej}, and further discussed in \cite{Preitschopf:1997gi,vanNieuwenhuizen:2004rh,Ortin:2004ms}. In their formulation, the supergravity is realized as a gauge theory of the corresponding supergroup $OSp(1|4;{\rm R})$. However, compared with the improved version for the pure Einstein gravity \cite{Stelle:1979aj,Wilczek:1998ea,Wang:2018tyq}, the disadvantage is that the covariance under all symmetries is not manifest in their construction. In this work, we aim to develop a covariant formulation of $N=1$ AdS$_{4}$ supergravity based on the AdS$^{4|4}$ bundle
\begin{eqnarray}
{\rm AdS}^{4|4} \longrightarrow \SE \overset{\pi}{\longrightarrow} \cM_{4}
%:\{\Vy^{\halpha},\Vtheta^{\hA}\}\to \{\Vy^{\halpha},x^{\mu}\} \overset{\pi}{\to} \{x^{\mu}\}
\end{eqnarray}
where the fiber space  AdS$^{4|4}$ is the rigid $N=1$ AdS$_{4}$ superspace \cite{Keck:1974se,Zumino:1977av,Ivanov:1980vb}
\begin{eqnarray}
{\rm AdS}^{4|4}=\frac{OSp(1|4;{\rm R})}{SO(1,3)}\,.
\end{eqnarray}
A different approach towards the gauge covariant construction of supergravity is the D'Auria-Fr\'{a} group-geometric formulation \cite{Neeman:1978zvv,DAdda:1980axn,DAuria:1980cmy,Regge:1983uu, Castellani:1981um,Castellani:1992sv,Castellani:2018zey} which is established in the language of soft supergroup manifold.

In Sec.II, based on the fact that the fermionic coordinates in AdS$^{4|4}$ form a faithful representation of supersymmetry, we analyse the linear representations of $OSp(1|4;{\rm R})$ systematically.  After embedding the superspace bosonic coordinates into the supervector linear representation, the supersymmetry-covariant construction of N=1 AdS$_4$ supergravity is established in Sec.III. Our notations for spinors are summarized in the appendix.
\section{Linear representations of N=1 AdS$_4$ supersymmetry}

\vspace{2ex}
\noindent \underline{\bf{N=1 AdS$_4$ superspace}}
\vspace{2ex}

The N=1 AdS$_4$ superspace AdS$^{4|4}$ is characterized by the supercoordiates $\{\Vy^{\halpha},\Vtheta^{\hA}\}$.
The bosonic coordinates $\{\Vy^{\halpha}\}$ form a vector of $SO(2,3)$ and satisfy the constraint
\begin{eqnarray}
\heta_{\halpha\hbeta}\Vy^{\halpha}\Vy^{\hbeta}=-\ell^2
\end{eqnarray}
where $\ell$ is the AdS radius.
The fermionic coordinates $\{\Vtheta^{\hA}\}$ form a Majorana spinor of $SO(2,3)$. The fermionic part of N=1 AdS$_4$ supersymmetry $OSp(1|4;{\rm R})$ can be realized by the non-linear super coordinate transformations
\begin{eqnarray}\label{dely}
  \delta \Vy^{\halpha} &=& \tfrac{\ii}{2\ell}\left[1-\tfrac{\ii}{4\ell}(\bar\Vtheta\Vtheta)-\tfrac1{8\ell^2}(\bar\Vtheta\Vtheta)^2\right]\Vy_{\hbeta}(\bar\Vxi \hgamma^{\hbeta\halpha}\Vtheta)\,,
\\\label{deltheta} \delta \Vtheta^{\hA}  &=& \left[1+\tfrac{\ii}{2\ell}(\bar\Vtheta\Vtheta) +\tfrac1{8\ell^2}(\bar\Vtheta\Vtheta)^2\right]\Vxi^{\hA}\,,
\end{eqnarray}
where the Majorana spinor $\Vxi^{\hA}$ is the infinitesimal fermionic translation parameter.
The original results in the literature \cite{Keck:1974se,Zumino:1977av,Ivanov:1980vb} can be recovered by redefining the fermionic coordinates as
\begin{eqnarray}
\theta^{\hA}=\left[1-\tfrac{\ii}{6\ell}(\bar\Vtheta\Vtheta)-\tfrac1{12\ell^2}(\bar\Vtheta\Vtheta)^2 \right] \Vtheta^{\hA}\,.
\end{eqnarray}

Under the fermionic translation (\ref{deltheta}), the scalar quantity
\begin{eqnarray}\label{hVw}
\hVw=\ell+\tfrac{\ii}2(\bar\Vtheta\Vtheta)+\tfrac1{8\ell}(\bar\Vtheta\Vtheta)^2\,,
\end{eqnarray}
transforms simply as
\begin{eqnarray}
\delta\hVw=\ii(\bar\Vxi\Vtheta)\,.
\end{eqnarray}
Including $\hVw$ as an additional auxiliary coordinate in the superspace, the fermionic translation can be rewritten as
\begin{eqnarray}\label{SUSYtrans}
\left(\begin{array}{c}
  \delta \Vy^{\halpha}   \\
  \delta \Vtheta^{\hA} \\
     \delta \hVw               \\
  \end{array}
 \right)=\left(\begin{array}{c}
   \ii (\ell+\hVw)^{-1}\Vy_{\hbeta}(\bar\Vxi \hgamma^{\hbeta\halpha}\Vtheta)  \\
\ell^{-1}\hVw\Vxi^{\hA}   \\
     \ii(\bar\Vxi\Vtheta)\\
  \end{array}
 \right)\,.
\end{eqnarray}
Inversely, if one takes (\ref{SUSYtrans}) as the starting point, the consistency of the supersymmetry algebra implies the constraint
\begin{eqnarray}\label{constr}
\hVw^2-\ii\ell  (\bar\Vtheta\Vtheta)=\ell^2\,.
\end{eqnarray}
Of course, there are two solutions for $\hVw$ in (\ref{constr}). One of the solutions is just (\ref{hVw}), and the other solution will lead to $\ell+\hVw=0$ at $(\bar\Vtheta\Vtheta)=0$. To avoid the singularity of $(\ell+\hVw)^{-1}$ in (\ref{SUSYtrans}) at $(\bar\Vtheta\Vtheta)=0$, the 2nd solution should be ruled out.

\vspace{2ex}
\noindent \underline{\bf{Superspinor representation}}
\vspace{2ex}

Applying the fermionic translation (\ref{deltheta}) twice, the commutator gives rise to the $SO(2,3)$ rotations on $\Vtheta^{\hA}$
\begin{eqnarray}
[\delta_{\oII},\delta_{\oI}] \Vtheta^{\hA}
=-\tfrac{\ii}{4\ell}
(\bar\Vxi_{\oI} \hgamma_{\halpha_1\halpha_2}\Vxi_{\oII})(\hgamma^{\halpha_1\halpha_2}\Vtheta)^{\hA}\,.
\end{eqnarray}
Therefore, the fermionic coordinates form a faithful non-linear representation of the corresponding supersymmetry algebra. In terms of the redundant coordinates, the superspinor
\begin{eqnarray}
\Stheta^{\ShA}=\left(
  \begin{array}{c}
    \Vtheta^{\hA}  \\
    \hVw  \\
  \end{array}
\right)
\end{eqnarray}
gives rise to a 5-dimensional faithful irreducible linear representation of N=1 AdS$_4$ supersymmetry
\begin{eqnarray}
\delta\Stheta^{\ShA}=\Su^{\ShA}{}_{\ShB}\Stheta^{\ShB}\,,~~~~~~~~\Su^{\ShA}{}_{\ShB}=\left(\begin{array}{cc}
0 & \ell^{-1}\Vxi^{\hA}\\
   \ii\bar\Vxi_{\hB} & 0 \\
  \end{array}
 \right)\,.
\end{eqnarray}

In the superspinor space, an important quantity is the superspinor metric
\begin{eqnarray}
\SC_{\ShA\ShB}=\left(
  \begin{array}{cc}
    -\ii\ell \hC_{\hA\hB} & 0\\
    0 & 1\\
  \end{array}
\right)\,,
\end{eqnarray}
where $\hC$ is the $SO(2,3)$ charge conjugation operator. It is symmetric under the super-permutation of indices
\begin{eqnarray}
\SC_{\ShA\ShB}=(-1)^{\ff[\ShA]\ff[\ShB]}\SC_{\ShB\ShA}
\end{eqnarray}
where $\ff[\ShA]$ denotes the fermion number.
One can easily check that $\SC_{\ShA\ShB}$ is an invariant super tensor, satisfying
\begin{eqnarray}
\delta\SC_{\ShA\ShB}=-(-1)^{\ff[\ShB](\ff[\ShA_1]+\ff[\ShA])}\SC_{\ShA_1\ShB}\Su^{\ShA_1}{}_{\ShA} -\SC_{\ShA\ShB_1}\Su^{\ShB_1}{}_{\ShB}=0\,.
\end{eqnarray}
Thus we can define the super co-spinor
\begin{eqnarray}
\bar\Stheta_{\ShA}=\SC_{\ShA\ShB}\Stheta^{\ShB}=\left(
  \begin{array}{cc}
    -\ii\ell \bar\Vtheta_{\hA} & \hVw
  \end{array}
\right)\,,
\end{eqnarray}
which transforms as
\begin{eqnarray}
\delta\bar\Stheta_{\ShA}=\bar\Stheta_{\ShB}\Su^{\ShB}{}_{\ShA}\,.
\end{eqnarray}
Obviously, the inner product
\begin{eqnarray}
(\bar\Stheta\Stheta)=\bar\Stheta_{\ShA}\Stheta^{\ShA}=-\ii\ell (\bar\Vtheta\Vtheta)+\hVw^2\,,
\end{eqnarray}
and thus the constraint (\ref{constr}) are invariant under the supersymmetry transformation.

Correspondingly, the right inverse superspinor metric $\SC^{\ShA\ShB}$ is defined by
\begin{eqnarray}
\SC_{\ShA\ShC}\SC^{\ShC\ShB}=\SI_{\ShA}{}^{\ShB}
\end{eqnarray}
where
\begin{eqnarray}
\SI_{\ShA}{}^{\ShB}=\left(
  \begin{array}{cc}
    -\delta_{\hA}{}^{\hB} & 0\\
    0 & 1\\
  \end{array}
\right)\,.
\end{eqnarray}
It implies that
\begin{eqnarray}
\SC^{\ShA\ShB}=\left(
  \begin{array}{cc}
    -\ii\ell^{-1} (\hC^{-1})^{\hA\hB} & 0\\
    0 & 1\\
  \end{array}
\right)
\end{eqnarray}
which is an invariant super tensor by construction.

\vspace{2ex}
\noindent \underline{\bf{Tensor products of superspinors}}
\vspace{2ex}

Now, one can easily generate an infinite number of linear representations by the tensor products of multiple superspinors. A tensor product representation is always reducible, and it can be decomposed into direct sum of irreducible representations which are characterized by Young diagrams.

In the simplest case, let us consider the $5\times5$ dimensional bi-superspinor representation $\Phi^{\ShA\ShB}$.
The bi-superspinor representation can be decomposed to the direct sum of 1-dimensional trace part
\begin{eqnarray}
\SC_{\ShB\ShA}\Phi^{\ShA\ShB}\,,
\end{eqnarray}
the $[(4+1)\times (4+1)]_{\Yboxdim{4pt}\yng(1,1)}=(4\times 1)+\frac{4\times(4+1)}2=14$ dimensional antisymmetric part
\begin{eqnarray}
\Phi_{\Yboxdim{4pt}\young(~,~)}^{\ShA\ShB}=\frac12\left[\Phi^{\ShA\ShB}-(-1)^{\ff[\ShA]\ff[\ShB]}\Phi^{\ShB\ShA}\right] =-(-1)^{\ff[\ShA]\ff[\ShB]}\Phi_{\Yboxdim{4pt}\yng(1,1)}^{\ShB\ShA}\,,
\end{eqnarray}
and the $[(4+1)\times (4+1)]_{\Yboxdim{4pt}\yng(2)}=(1\times 1)+(4\times 1)+\frac{4\times(4-1)}2-1=10$ dimensional symmetric traceless part
\begin{eqnarray}
\Phi_{\Yboxdim{4pt}\young(~~)}^{\ShA\ShB}=\frac12\left[\Phi^{\ShA\ShB}+(-1)^{\ff[\ShA]\ff[\ShB]}\Phi^{\ShB\ShA}\right] +\frac13\SC_{\ShB_1\ShA_1}\Phi^{\ShA_1\ShB_1}\SC^{\ShA\ShB} =(-1)^{\ff[\ShA]\ff[\ShB]}\Phi_{\Yboxdim{4pt}\yng(2)}^{\ShB\ShA}\,.
\end{eqnarray}

Especially, the $14$-dimensional antisymmetric representation is given by
\begin{eqnarray}
(\Phi_{\Yboxdim{4pt}\young(~,~)})^{\ShA}{}_{\ShB}&=&(-1)^{\ff[\ShA](\ff[\ShB]+\ff[\ShC])}\SC_{\ShB\ShC}(\Phi_{\Yboxdim{4pt}\young(~,~)})^{\ShA\ShC}=\left(
  \begin{array}{cc}
    \frac14 {\slashed s}^{\hA}{}_{\hB} & \ell^{-1}\kappa^{\hA}\\
    \ii \bar\kappa_{\hB} & 0 \\
  \end{array}
\right)\,,
\end{eqnarray}
where ${\slashed s}=s^{\halpha\hbeta}\hgamma_{\halpha\hbeta}$%\,, and the 2nd indices is the lowered by $\SC_{\ShB\ShC}$ for simplicity
.
In terms of the $SO(2,3)$ representations, $\Phi_{\Yboxdim{4pt}\young(~,~)}$ consists of an antisymmetric 2-tensor $s^{\halpha\hbeta}$ and a Majorana spinor $\kappa^{\hA}$. Obviously, $\{s^{\halpha\hbeta},\kappa^{\hA}\}$ form the adjoint representation for which the fermionic translation is given by
\begin{eqnarray}
\left(
  \begin{array}{c}
    \delta s^{\halpha\hbeta}  \\
    \delta \kappa^{\hA} \\
  \end{array}
\right)
=\left(
  \begin{array}{c}
    -\frac\ii{\ell}(\bar\Vxi\hgamma^{\halpha\hbeta}\kappa)  \\
    -\frac1{4}({\slashed s}\Vxi)^{\hA} \\
  \end{array}
\right)\,.
\end{eqnarray}
\vspace{2ex}
\noindent \underline{\bf{Supervector representation}}
\vspace{2ex}

On the other hand, the $10$-dimensional symmetric traceless representation is given by
\begin{eqnarray}
(\Phi_{\Yboxdim{4pt}\young(~~)})^{\ShA}{}_{\ShB}&=&\left(
  \begin{array}{cc}
    \frac{1}{4}\phi\delta^{\hA}{}_{\hB} +\frac{1}{4}\slashed v^{\hA}{}_{\hB} & \ell^{-1}\bar\zeta^{\hA}\\
    -\ii \zeta_{\hB} & \phi \\
  \end{array}
\right)\,,
\end{eqnarray}
where $\slashed v=v^{\halpha}\hgamma_{\halpha}$.
In terms of the $SO(2,3)$ representations, it consists of a vector $v^{\halpha}$, a Majorana spinor $\zeta^{\hA}$ and a scalar $\phi$. Thus we shall name it as the supervector representation. In terms of  ${\mathbbm v}^{\Shalpha}=\{v^{\halpha},\zeta^{\hA},\phi\}$, the fermionic translation is expressed as
\begin{eqnarray}
\delta {\mathbbm v}^{\Shalpha}=\Su^{\Shalpha}{}_{\Shbeta}{\mathbbm v}^{\Shbeta}\,,~~~~~~
\Su^{\Shalpha}{}_{\Shbeta}=\left(
  \begin{array}{ccc}
   0 & 2\ii\ell^{-1}(\bar\Vxi\hgamma^{\halpha})_{\hB} & 0\\
    -\frac{1}{4} (\hgamma_{\hbeta}\Vxi)^{\hA} & 0 & \frac34 \Vxi^{\hA} \\
     0 & 2\ii\ell^{-1}\bar\Vxi_{\hB} & 0\\
  \end{array}
\right)\,.
\end{eqnarray}
In the supervector space, the symmetric super 2-tensor
\begin{eqnarray}
\Seta_{\Shalpha\Shbeta}=
\left(
  \begin{array}{ccc}
  \heta_{\halpha\hbeta} & 0 & 0\\
   0 & 8\ii\ell^{-1} \hC_{\hA\hB} & 0\\
   0 & 0 & -3\\
  \end{array}
\right)=(-1)^{\ff[\Shalpha]\ff[\Shbeta]}\Seta_{\Shbeta\Shalpha}
\end{eqnarray}
is an invariant super tensor, satisfying
\begin{eqnarray}
\delta\Seta_{\Shalpha\Shbeta}=-(-1)^{\ff[\Shbeta](\ff[\Shalpha_1]+\ff[\Shalpha])}\Seta_{\Shalpha_1\Shbeta}\Su^{\Shalpha_1}{}_{\Shalpha} -\Seta_{\Shalpha\Shbeta_1}\Su^{\Shbeta_1}{}_{\Shbeta}=0\,.
\end{eqnarray}

Since the superspace bosonic coordinates $\{\Vy^{\halpha}\}$ form a $SO(2,3)$ vector, it is natural to ask whether $\{\Vy^{\halpha}\}$ could be embedded in the supervector representation. After some efforts, we find that the answer is yes. The corresponding supervector $\Sy^{\Shalpha}$ is given by
\begin{eqnarray}
\Sy^{\Shalpha}=\left(
  \begin{array}{c}
    4(1+\frac1{2\ell}\Vw)^{-1}\Vy^{\halpha} \\
    -(1+\frac1{2\ell}\Vw)^{-2}(\slashed\Vy\Vtheta)^{\hA} \\
    -\frac{\ii}{\ell}(\bar\Vtheta\slashed\Vy\Vtheta) \\
  \end{array}
\right)=\left(
  \begin{array}{c}
    4\left[1-\frac{\ii}{4\ell}(\bar\Vtheta\Vtheta)-\frac{1}{8\ell^2}(\bar\Vtheta\Vtheta)^2\right]\Vy^{\halpha} \\
    -\left[1-\frac{\ii}{2\ell}(\bar\Vtheta\Vtheta)\right](\slashed\Vy\Vtheta)^{\hA} \\
    -\frac{\ii}{\ell}(\bar\Vtheta\slashed\Vy\Vtheta) \\
  \end{array}
\right)\,,
\end{eqnarray}
where $\Vw=\hVw-\ell$. In terms of the bi-superspinor notation, the supervector $\Sy^{\Shalpha}$ becomes
\begin{eqnarray}
\Sy^{\ShA}{}_{\ShB}=\left(
  \begin{array}{cc}
    -\tfrac{\ii}{4\ell}(\bar\Vtheta\slashed\Vy\Vtheta)\delta^{\hA}{}_{\hB} +\left[1-\frac{\ii}{4\ell}(\bar\Vtheta\Vtheta)-\tfrac{1}{8\ell^2}(\bar\Vtheta\Vtheta)^2\right]\slashed\Vy^{\hA}{}_{\hB} & -\ell^{-1}\!\left[1-\frac{\ii}{2\ell}(\bar\Vtheta\Vtheta)\right](\slashed\Vy\Vtheta)^{\hA}\\
    \ii\left[1-\frac{\ii}{2\ell}(\bar\Vtheta\Vtheta)\right](\bar\Vtheta\slashed\Vy)_{\hB} & -\ii\ell^{-1}(\bar\Vtheta\slashed\Vy\Vtheta)\\
  \end{array}
\right)\,.
\end{eqnarray}
It is noticed that
\begin{eqnarray}
\Sy^{\ShA}{}{}_{\ShB}\Stheta^{\ShB}&=&0\,,
\\\label{yy}
\Sy^{\ShA}{}{}_{\ShC}\Sy^{\ShC}{}_{\ShB}&=&\Stheta^{\ShA}\bar\Stheta_{\ShB}-\ell^2
    \delta^{\ShA}{}_{\ShB}\,.
\end{eqnarray}
The supertrace of (\ref{yy})
\begin{eqnarray}
{\rm Str}(\Sy\Sy)&=&\bar\Stheta_{\ShA}\Stheta^{\ShA}-\ell^2
    \SI_{\ShA}{}^{\ShA}%=\ell^2+3\ell^2
    =4\ell^2\,,
\end{eqnarray}
is equivalent to the norm in the supervector notation
\begin{eqnarray}
\Seta_{\Shalpha\Shbeta}\Sy^{\Shbeta}\Sy^{\Shalpha}=-16\ell^2\,.
\end{eqnarray}

\section{N=1 AdS$_4$ supergravity}
%\vspace{2ex}
%\noindent \underline{\bf{Einstein gravity}}
%\vspace{2ex}

\vspace{2ex}
\noindent \underline{\bf{Gauging the supersymmetry}}
\vspace{2ex}

In the supergravity theory, the AdS$_4$ supersymmetry $OSp(1|4;{\rm R})$ becomes the local gauge symmetry on the base manifold $\cM_4$. Mathematically, we are considering fiber bundles for which each fiber is a representation space of supersymmetry.  The spacetime superfields on $\cM_4$ are sections of the super bundles. E.g., a section of the superspinor bundle is
\begin{eqnarray}
\Stheta^{\ShA}{}_{\ShB}=\STheta^{\ShA}{}_{\ShB}(x)~~~&\Leftrightarrow&~~~ \Vtheta^{\hA}=\Theta^{\hA}(x)\,,
\end{eqnarray}
and a section of the supervector bundle is
\begin{eqnarray}
\Sy^{\ShA}{}_{\ShB}=\SY^{\ShA}{}_{\ShB}(x)~~~&\Leftrightarrow&~~~ \{\Vy^{\halpha}=Y^{\halpha}(x)\,,~~\Vtheta^{\hA}=\Theta^{\hA}(x)\}\,.
\end{eqnarray}
Via the local supersymmetry transformation
\begin{eqnarray}
\Su^{\ShA}{}_{\ShB}(x)=\left(\begin{array}{cc}
\frac14\slashed \Bu^{\hA}{}_{\hB}(x) & \ell^{-1}\Vxi^{\hA}(x)\\
   \ii\bar\Vxi_{\hB}(x) & 0 \\
  \end{array}
 \right)\,,
\end{eqnarray}
one can always get to the super Einstein gauge
\begin{eqnarray}
\{Y^{\hblt}(x)=\ell,~~~Y^{\ha}(x)=0\,,~~~\Theta^{\hA}(x)=0\}\,.
\end{eqnarray}
%a section of the superspace bundle $\{\Vy^{\halpha},\Vtheta^{\hA};x^{\mu}\}$ is given by
%\begin{eqnarray}
%\Vy^{\halpha}=Y^{\halpha}(x)\,,~~~~~~~~\Vtheta^{\hA}=\Theta^{\hA}(x)\,.
%\end{eqnarray}

To define the covariant derivatives compatible with the local supersymmetry transformations, we further introduce the corresponding $OSp(1|4;{\rm R})$  gauge field 1-form on $\cM_4$
\begin{eqnarray}
\SA^{\ShA}{}_{\ShB}(x)=\left(\begin{array}{cc}
\frac14\slashed A{}^{\hA}{}_{\hB}(x) & \ell^{-1}\psi^{\hA}(x)\\
   \ii\bar\psi_{\hB}(x) & 0 \\
  \end{array}
 \right)
\end{eqnarray}
which is valued in the adjoint representation of local $OSp(1|4;{\rm R})$. The $OSp(1|4;{\rm R})$  gauge field $\SA^{\ShA}{}_{\ShB}$ unifies the spacetime Vielbein
\begin{eqnarray}
e^{\ha}{}_{\hblt}(x)=\ell A^{\ha}{}_{\hblt}(x)=\ell \SA^{\ha}{}_{\hblt}(x)\,,
\end{eqnarray}
the spin connection
\begin{eqnarray}
\omega^{\ha}{}_{\hb}(x)=A^{\ha}{}_{\hb}(x)=\SA^{\ha}{}_{\hb}(x)\,,
\end{eqnarray}
as well as the gravitino field $\psi^{\hA}(x)$.
Under local supersymmetry transformation, the $OSp(1|4;{\rm R})$ gauge field transforms as
\begin{eqnarray}
\delta\SA^{\ShA}{}_{\ShB}=-\rd\Su^{\ShA}{}_{\ShB}-\SA^{\ShA}{}_{\ShC}\Su^{\ShC}{}_{\ShB} +\Su^{\ShA}{}_{\ShC}\SA^{\ShC}{}_{\ShB}=-\SD\Su^{\ShA}{}_{\ShB}\,,
\end{eqnarray}
such that
\begin{eqnarray}
\SD\SY^{\ShA}{}_{\ShB}=\rd\SY^{\ShA}{}_{\ShB}+\SA^{\ShA}{}_{\ShC}\SY^{\ShC}{}_{\ShB} -\SY^{\ShA}{}_{\ShC}\SA^{\ShC}{}_{\ShB}
\end{eqnarray}
also forms the super vector representation.
Especially, the fermionic translations on the components are given by
\begin{eqnarray}
\delta e^{\ha}&=&-\ii(\bar\Vxi\hgamma^{\ha}{}_{\hblt}\psi)\,,~~~~~~~~ \delta\omega^{\ha}{}_{\hb}=-\frac\ii{\ell}(\bar\Vxi\hgamma^{\ha}{}_{\hb}\psi)\,,
~~~~~~~~~~
\delta\psi^{\hA}=-\rD\Vxi^{\hA}\,,%=-\underline{\rD}\Vxi^{\hA}+
\end{eqnarray}
where $\rD=\rd+A$ denotes the $SO(2,3)$ covariant derivative. %, ??and $\underline{\rD}$ is the $SO(1,3)$ covariant derivative $\underline{\rD}=\rd+\omega$

The commutator $[\SD,\SD]$ gives rise to 2-form gauge field strength of supersymmetry
\begin{eqnarray}
\SF^{\ShA}{}_{\ShB}&=&\rd\SA^{\ShA}{}_{\ShB}+\SA^{\ShA}{}_{\ShC}\wedge \SA^{\ShC}{}_{\ShB}
\cr&=&\left(\begin{array}{cc}
\frac14(\slashed \BF)^{\hA}{}_{\hB}-\frac{\ii}8\ell^{-1}(\bar\psi\wedge\hgamma_{\halpha_1\halpha_2}\psi)(\hgamma^{\halpha_1\halpha_2})^{\hA}{}_{\hB} & \ell^{-1}\rD\psi^{\hA}\\
   \ii\rD\bar\psi_{\hB} & 0 \\
  \end{array}
 \right)
=\left(\begin{array}{cc}
\frac14(\slashed\BSF)^{\hA}{}_{\hB} & \ell^{-1}\rD\psi^{\hA}\\
   \ii\rD\bar\psi_{\hB} & 0 \\
  \end{array}
 \right)
\end{eqnarray}
where $F=[\rD,\rD]$ is the $SO(2,3)$ field strength and
\begin{eqnarray}
\BSF^{\halpha_1\halpha_2}= \BF^{\halpha_1\halpha_2}-\frac{\ii}2\ell^{-1}(\bar\psi\wedge\hgamma^{\halpha_1\halpha_2}\psi)\,.
\end{eqnarray}
By construction, $\SF^{\ShA}{}_{\ShB}$ forms the adjoint representation and transforms as a tensor under local supersymmetry
\begin{eqnarray}
\delta\SF^{\ShA}{}_{\ShB}=\Su^{\ShA}{}_{\ShC}\SF^{\ShC}{}_{\ShB} -\SF^{\ShA}{}_{\ShC}\Su^{\ShC}{}_{\ShB}\,.
\end{eqnarray}

%One can easily construct new super tensor fields out of $\STheta$, $\SY$, $\SF$ and their covariant derivatives, e.g.,  the bi-superspinor with a free parameter $\lambda$
%\begin{eqnarray}
%\SSF^{\ShA}{}_{\ShB}=
%\SF^{\ShA}{}_{\ShB} -\frac{\lambda}{\ell^2}[\SF^{\ShA}{}_{\ShC}\STheta^{\ShC}\bar\STheta_{\ShB} +\Stheta^{\ShA}\bar\STheta_{\ShC}\SF^{\ShC}{}_{\ShB}] \,.
%\end{eqnarray}
%The quantity $\SSF^{\ShA}{}_{\ShB}$ will be useful in constructing the N=1 AdS$_4$ supergravity action.

\vspace{2ex}
\noindent \underline{\bf{Supergravity action}}
\vspace{2ex}

To recover the N=1 AdS$_4$ supergravity action, we introduce the $\lambda$-dependent bi-superspinor
\begin{eqnarray}
\SSF^{\ShA}{}_{\ShB}=
\SF^{\ShA}{}_{\ShB} -\frac{\lambda}{\ell^2}[\SF^{\ShA}{}_{\ShC}\STheta^{\ShC}\bar\STheta_{\ShB} +\STheta^{\ShA}\bar\STheta_{\ShC}\SF^{\ShC}{}_{\ShB}]
\end{eqnarray}
which is in the adjoint representation. The value of the parameter $\lambda$ shall be decided later on. Analogy with the MacDowell-Mansouri-Stelle-West action, we propose a Lagrangian density
\begin{eqnarray}
I_4=\STr(\SSF\wedge\SSF\,\SY)=\STr(\SF\wedge\SF\,\SY)+\frac{\lambda(\lambda-2)}{\ell^2}(\bar\STheta\SF\wedge\SY\SF\STheta)\,
\end{eqnarray}
which is manifestly supersymmetry-invariant in our construction.

Expanding $I_4$ by the powers of $\Theta(x)$, we get
\begin{eqnarray}
I_4&=&\frac{1}{4}\epsilon_{\halpha_1\halpha_2\halpha_3\halpha_4\halpha_5} \BSF^{\halpha_1\halpha_2}\wedge\BSF^{\halpha_3\halpha_4}Y^{\halpha_5}
+\frac{\ii(1-\lambda)^2}{\ell}(\rD\bar\psi\slashed Y\wedge\rD\psi) \cr&&+\frac{\ii\lambda(2-\lambda)}{2\ell}(\bar\Theta\slashed \BSF\slashed Y\wedge\rD\psi)
+\frac{\ii }{2\ell}(\bar\Theta\slashed Y\slashed \BSF\wedge\rD\psi)+O(\Theta^2)\,.
\end{eqnarray}
At the point of super Einstein gauge
\begin{eqnarray}\label{Einp}
\{Y^{\hblt}(x)=\ell,~~~Y^{\ha}(x)=0\,,~~~\Theta^{\hA}(x)=0\}\,,
\end{eqnarray}
the Lagrangian density $I_4$ becomes
\begin{eqnarray}
I_4^{(0)}&=&\frac{1}{4}\epsilon_{\halpha_1\halpha_2\halpha_3\halpha_4\halpha_5} \BSF^{\halpha_1\halpha_2}\wedge\BSF^{\halpha_3\halpha_4}Y^{\halpha_5}
+\frac{\ii(1-\lambda)^2}{\ell}(\rD\bar\psi\slashed Y\wedge\rD\psi)
\cr&=&\frac{\ell}{4}\epsilon_{\ha_1\ha_2\ha_3\ha_4} \BSF^{\ha_1\ha_2}\wedge\BSF^{\ha_3\ha_4} +\ii(1-\lambda)^2(\rD\bar\psi\gamma\wedge\rD\psi)
\,,
\end{eqnarray}
which is in the form presented in \cite{MacDowell:1977jt,Mansouri:1977ej,Preitschopf:1997gi,vanNieuwenhuizen:2004rh,Ortin:2004ms}. Meanwhile,
the infinitesimal fermionic translation becomes
\begin{eqnarray}\label{delta0}
\delta Y^{\halpha} &=& 0\,,~~~~~~~~~~~~~~~~~~\delta \Theta^{\hA}=\Vxi^{\hA}\,,~~~~~~~~
\cr\delta e^{\ha}&=&-\ii(\bar\Vxi\hgamma^{\ha}{}_{\hblt}\psi)\,,~~~~~~~~ \delta\omega^{\ha}{}_{\hb}=-\frac\ii{\ell}(\bar\Vxi\hgamma^{\ha}{}_{\hb}\psi)\,,
~~~~~~~~~~
\delta\psi^{\hA}=-\rD\Vxi^{\hA}\,.
\end{eqnarray}
Under the reduced fermionic translation (\ref{delta0}), the transformation of $I_4^{(0)}$ is totally fixed by the original supersymmetry invariance of $I_4$ at the point (\ref{Einp}). The result is simply
\begin{eqnarray}
-\delta I_4^{(0)}&=& \frac{\ii\lambda(2-\lambda)}{2\ell}(\bar\Vxi\slashed \BSF\slashed Y\wedge\rD\psi)
+\frac{\ii }{2\ell}(\bar\Vxi\slashed Y\slashed \BSF\wedge\rD\psi)\,,
\end{eqnarray}
which can be directly read off from the $O(\Theta)$ term of $I_4$.

Especially, when $\lambda=1\pm\sqrt2$, we have
\begin{eqnarray}\label{S_0}
S_4^{(0)}&=& \int_{\cM_4}\left[\frac{1}{4}\epsilon_{\halpha_1\halpha_2\halpha_3\halpha_4\halpha_5} \BSF^{\halpha_1\halpha_2}\wedge\BSF^{\halpha_3\halpha_4}Y^{\halpha_5}+\frac{2\ii }{\ell}(\rD\bar\psi\slashed Y\wedge\rD\psi)\right]\,,
%\left[\frac{1}{4}\epsilon_{\ha_1\ha_2\ha_3\ha_4} \BSF^{\ha_1\ha_2}\wedge\BSF^{\ha_3\ha_4} +2\ii (\rD\bar\psi\gamma\wedge\rD\psi)\right]
\end{eqnarray}
as well as
\begin{eqnarray}
\delta S_4^{(0)}&=& %\frac{\ii }{2\ell}(\bar\Vxi[\slashed\BSF, \slashed\Vy ]\wedge\rD\psi)=
\frac{\ii }{\ell} \int_{\cM_4}\BSF^{\halpha}{}_{\hbeta}Y^{\hbeta}\wedge(\bar\Vxi\hgamma_{\halpha}\rD\psi)
%=\frac{1 }{\ell} \int_{\cM_4}\BSF^{\ha}{}_{\hblt}(\bar\Vxi\gamma\gamma_{\ha}\wedge\rD\psi)
\,.
\end{eqnarray}
Thus the reduced fermionic translation (\ref{delta0})  shall vanish for (\ref{S_0}) by satisfying the super torsion free condition
\begin{eqnarray}\label{tf}
\BSF^{\halpha}{}_{\hbeta}Y^{\hbeta}=\ell\BSF^{\ha}{}_{\hblt} =0\,.
\end{eqnarray}
On the other hand, the super torsion free condition (\ref{tf}) is equivalent to the equation of motion from varying $\omega^{\ha}{}_{\hb}$ in the action (\ref{S_0})
\begin{eqnarray}
\frac{\delta S_4^{(0)}}{\delta\omega^{\ha}{}_{\hb}}=0~~~~~~~~&\Leftrightarrow&~~~~~~~~ \BSF^{\ha}{}_{\hblt}=0\,.
\end{eqnarray}
Therefore, in the 2nd order formalism where the spin connection $\omega^{\ha}{}_{\hb}$ is integrated out, the action (\ref{S_0}) is invariant under the transformation
\begin{eqnarray}
\delta e^{\ha}&=&-\ii(\bar\Vxi\hgamma^{\ha}{}_{\hblt}\psi)\,,~~~~~~~~~~~~
\delta\psi^{\hA}=-\rD\Vxi^{\hA}\,.
\end{eqnarray}
This exactly reproduces the 1.5-order argument \cite{MacDowell:1977jt,Mansouri:1977ej,Preitschopf:1997gi,vanNieuwenhuizen:2004rh,Ortin:2004ms} for the supersymmetry invariance of %the on-shell formalism for
N=1 AdS$_4$ supergravity.

\section{Summary}
Along the lines of \cite{MacDowell:1977jt,Mansouri:1977ej,Stelle:1979aj,Wilczek:1998ea,Wang:2018tyq, Preitschopf:1997gi,vanNieuwenhuizen:2004rh}, a manifestly supersymmetry-covariant approach for constructing N=1 AdS$_4$ supergravity is established based on a systematical description of the finite dimensional linear representations of the supersymmetry. The crucial step of this construction is the non-linear embedding of AdS superspace coordinates into certain linear representation of the supersymmetry group.

As the subsequent tasks, we hope to extend this analysis to $N>1$ as well as different dimensions. For generic dimensions, it is more natural to consider the super counterpart of the action (\ref{VW}) instead of the MacDowell-Mansouri-Stelle-West type of action. This might lead to a simple and universal description of supergravity in the superspace language.

\appendix
\setcounter{equation}{0}
\setcounter{subsection}{0}
\renewcommand{\theequation}{A.\arabic{equation}}
\renewcommand{\thesubsection}{A.\arabic{subsection}}
\section{Convention for spinors}
The $SO(1,3)$ Clifford algebra is  
\begin{eqnarray}
\{\gamma^{\ha},\gamma^{\hb}\}=2\eta^{\ha\hb}\bm1\,,~~~~~\eta={\rm diag}\{-1,1,1,1\}
\end{eqnarray}
where $\ha,\hb=\hat0,\hat1,\hat2,\hat3$ are $SO(1,3)$ vector indices. The gamma matrices $(\gamma^{\ha})^{\hA}{}_{\hB}$ acts on the 4-components spinor $\Vxi^{\hA}$.
The chirality operator $\gamma$ is given by
\begin{eqnarray}
\gamma=-\ii\gamma^{\hat0}\gamma^{\hat1}\gamma^{\hat2}\gamma^{\hat3}
\end{eqnarray}
which satisfies that
\begin{eqnarray}
\gamma^2=\bm1\,.
\end{eqnarray}
The $SO(1,3)$ charge conjugation operator $C$ satisfies that
\begin{eqnarray}
C\gamma^{\ha}C^{-1}=-(\gamma^{\ha})^{T}\,,~~~~~~~~~~~~C^T=-C\,.
\end{eqnarray}
Thus
\begin{eqnarray}
(C\gamma^{\ha_1\cdots\ha_n})^T&=&(-1)^{\frac{(n-2)(n-1)}2}C\gamma^{\ha_1\cdots\ha_n}\,.
\end{eqnarray}
A $SO(1,3)$ co-spinor $\bar\Vxi_{\hA}$ can be defined as
\begin{eqnarray}
\bar\Vxi=\Vxi^{\dagger}\gamma^{\hat0}\,.
\end{eqnarray}
The Majorana spinor is defined by
\begin{eqnarray}
\bar\Vxi_{\hA}=C_{\hA\hB}\Vxi^{\hB}\,.
\end{eqnarray}

It is easy to realize the $SO(2,3)$ Clifford algebra
\begin{eqnarray}
\{\hgamma^{\halpha},\hgamma^{\hbeta}\}=2\heta^{\halpha\hbeta}\bm1\,,~~~~~\heta={\rm diag}\{-1,-1,1,1,1\}
\end{eqnarray}
by the identification
\begin{eqnarray}
\hgamma^{\halpha}=\{\hgamma^{\hblt}=\ii\gamma,\hgamma^{\ha}=\ii\gamma\gamma^{\ha} \}
\end{eqnarray}
where $\halpha,\hbeta=\hblt,\hat0,\hat1,\hat2,\hat3$ are $SO(2,3)$ vector indices. Thus the $SO(2,3)$ spinor space is the same as the $SO(1,3)$ spinor space. The  $SO(2,3)$ charge conjugation operator $\hC$ is just the $SO(1,3)$ one
\begin{eqnarray}
\hC=C\,.
\end{eqnarray}
Thus
\begin{eqnarray}
\hC^T=-\hC\,,~~~~~~(\hC\hgamma^{\halpha})^T=-\hC\hgamma^{\halpha}\,,~~~~~~ (\hC\hgamma^{\halpha_1\halpha_2})^T=\hC\hgamma^{\halpha_1\halpha_2}\,.
\end{eqnarray}
A $SO(2,3)$ co-spinor $\bar\Vxi_{\hA}$ is defined by
\begin{eqnarray}
\bar\Vxi=\Vxi^{\dagger}\hgamma^{\hat0}\hgamma^{\hblt}=\Vxi^{\dagger}\gamma^{\hat0}\,,
\end{eqnarray}
which is equivalent to the $SO(1,3)$ case. Thus a $SO(1,3)$ Majorana spinor is also the $SO(2,3)$ Majorana spinor.
The inner product between two spinors is denoted simply as
\begin{eqnarray}
(\bar\Vxi\Vtheta)=\bar\Vxi_{\hA}\Vtheta^{\hA}\,.
\end{eqnarray}

\vspace*{3.0ex}
\begin{acknowledgments}
\paragraph*{Acknowledgments.}
The authors thank Bo-Han Li and Hong L\"u for useful conversations.
This work is supported by National Natural Science Foundation of China(Grants No. 12275217, No. 12247103).
\end{acknowledgments}

\bibliographystyle{unsrturl}
%\bibliography{AdS4SUSY}

%\onecolumngrid

\end{document}